# A quantitative measure, mechanism and attractor for self-organization in networked complex systems

Georgi Yordanov Georgiev

Department of Natural Sciences – Physics and Astronomy, Assumption College,
500 Salisbury St, Worcester MA, 01609, United States of America
ggeorgie@assumption.edu, georgi@alumni.tufts.edu

**Abstract.** Quantity of organization in complex networks here is measured as the inverse of the average sum of physical actions of all elements per unit motion multiplied by the Planck's constant. The meaning of quantity of organization is the inverse of the number of quanta of action per one unit motion of an element. This definition can be applied to the organization of any complex system. Systems self-organize to decrease the average action per element per unit motion. This lowest action state is the attractor for the continuous self-organization and evolution of a dynamical complex system. Constraints increase this average action and constraint minimization by the elements is a basic mechanism for action minimization. Increase of quantity of elements in a network, leads to faster constraint minimization through grouping, decrease of average action per element and motion and therefore accelerated rate of self-organization. Progressive development, as self-organization, is a process of minimization of action.

**Keywords:** network, self-organization, complex system, organization, quantitative measure, principle of least action, principle of stationary action, attractor, progressive development, acceleration

## 1 Introduction

### 1.1 Motivation

To define quantitatively self-organization in complex networked systems a measure for organization is necessary [1]. Two systems should be numerically distinguishable by their degree of organization and their rate of self-organization. What one quantity can measure the degree of self-organization in all complex systems? To answer this question we turn to established science principles and ask: What single principle can explain the largest number of science phenomena? It turns out that this is the principle of least (stationary) action. There is no more broad and fundamental principle in science than this, as it can be seen in the next section.

### 1.2 The principle of least action and its variations

Pierre de Maupertuis stated Law of the Least Action as a "universal principle from which all other principles naturally flow" [2]. Later Euler, Lagrange, Hamilton, Fermat, Einstein, and many others refined it and applied it to develop all areas of physics [3]. It was later generalized as a path integral formalism for quantum mechanics by Feynman [4]. Jacobi's form of the principle refers to the path of the system point in a curvilinear space characterized by the metric tensor [3]. The Hertz's principle of least curvature says that a particle tends to travel along the path with least curvature, if there are not external forces acting on it [3]. The Gauss Principle of least constraint where the motion of a system of interconnected material points is such as to minimize the constraint on the system is an alternative formulation of classical mechanics, using a differential variational principle [5]. Action is more general than energy and any law derived from the principle of least action is guaranteed to be self consistent [6]. All of the laws of motion and conservation in all branches of physics are derived from the principle of least action [6,7].

### 1.3 Applications to networks and complex systems

Scientists have applied the principle of least action to networks and complex systems. For example, it has been applied to network theory [8,9,10] and path integral approaches to stochastic processes and networks [11]. Samples of some other applications are by Annila and Salthe for natural selection [12] and Devezas for technological change [13]. Some of the other important measures and methods used in complex systems research are presented by Chaisson [14], Bar-Yam [15], Smart [16], Vidal [17] and Gershenson and Heylighen [18]. This list is not exhaustive. Some of these established measures use information, entropy or energy to describe complexity, while a fundamental quantity of physical action is used in this work to describe degree of organization through efficiency.

### 2    Principle of least action for a system of elements – an attractor

In a previous paper [1] we defined the natural state of an organized system as the one in which the variation of the sum of actions of all of the elements is zero. Here we define the principle of least action for n elements crossing m nodes as:

$$\delta \sum_{i=0}^{n} \sum_{j=0}^{m} I_{ij} = \delta \sum_{i=1}^{n} \sum_{j=1}^{m} \int_{t_1}^{t_2} L_{ij} dt = 0 \;. \tag{1}$$

Where $\delta$ is infinitesimally small variation in the action integral $I_{ij}$ of the j[th] crossings between the nodes (unit motion) of the i[th] element and $L_{ij}$ is the Lagrangian for that motion. *n* represents the number of elements in a system, m the number of motions and $t_1$ and $t_2$ are the initial and final times of each motion. $\sum_{i=0}^{n} \sum_{j=0}^{m} I_{ij}$ is the sum of all actions of all elements n for their motions m between nodes of a complex network. For example, a unit motion for electrons on a computer chip is the one necessary for one computation. For a computer network, such as internet, it is the transmission of one bit of information. In a chemical system it is the one for one chemical reaction. The state of zero variation of the total action for all motions is the one to which any system is naturally driven. Open systems never achieve this least action state because of the constant changes that occur in them, but are always tending toward it. In some respect one can consider this attractor state to be one of dynamical action equilibrium. Using the quantity of action one can measure how far the system is from this equilibrium and can distinguish between the organizations of two systems, both of which are equally close to equilibrium.

## 3 Physical Action as a quantitative measure for organization

In [1] we defined organization of a system as inversely proportional to the average sum of all actions. Here we expand this notion by defining organization, α, as inversely proportional to the average action per one element and one motion.

$$\alpha = \frac{hnm}{\sum_{i=0}^{n}\sum_{j=0}^{m} I_{ij}} \quad . \tag{2}$$

h is the Planck's constant. The meaning of organization is that it is inversely proportional to the number of quanta of action per one motion of one element in a system. This definition is for a system of identical elements or where elements can be approximated as identical. It is the efficiency of physical action. The time derivative of α is the rate of progressive development of a complex system.

We can expand this for a system of non-identical elements, which is the most general case. Let's write instead of α for each type of elements, its reciprocal, which is the number of quanta, q, for each of those elements to cross one edge in the network. Then:

$$\alpha = \frac{\sum_k n_k}{\sum_k q_k \; n_k} \tag{3}$$

Where $n_k$ is the k-th type of element, and $q_k$ is the average number of quanta per element and edge crossing for the k-th element. It can be verified by plugging sample numbers that the average number of quanta per element per edge crossing for the system is weighted by how many elements are of the type with particular efficiency. The sum of products of the number of elements of each type, times the average number of quanta per element and edge crossing for that type of element, divided by the sum of the number of elements of each type, gives the average number of quanta per element per edge crossing in the system, weighted by the number of elements of each type. The reciprocal of that is the total organization of the system, or α, but now given by the new equation 3.

If two systems with two different values of α compete, the one with the higher value will have advantage. Within each system, the elements with higher efficiency (lower average number of quanta per element and edge crossing) will win over the rest, in order to increase α for the system, as given by eq. 3. If the elements with lower efficiency can reorganize themselves, they will survive if they match the efficiency of the most efficient ones. For the purposes of calculating α, even if the elements in a system are different otherwise, and only their efficiencies are identical, they can be treated not as identical elements, but as different elements with identical efficiencies, in which case, eq. 3 reduces to eq. 2.

Here we acknowledge that efficiency takes into account the changing external conditions for the system. If a system has certain efficiency in given conditions, when they change, the constraints for the system will change as well, and the efficiency of the system will drop immediately, because the system has not yet had time to optimize the new constraints. At each change of the surrounding conditions for the system, its efficiency is generally decreased, and it starts a new process of constraint optimization to increase it again. If the system is small, the change can decrease efficiency so much as to destroy the system. If the system is at advanced stage of its development, it can minimize the new constraints much faster, therefore its robustness increases with level of organization.

# 4  Applications

## 4.1  One element and one constraint

Consider the simplest possible part of a network: one edge, two nodes and one element moving from node 1 to node 2. Let's consider case (I) when there is no constraint for the motion of the element. It crosses the path between nodes 1 and 2 along the shortest line – a geodesic. Now consider case (II) when there is one constraint placed between nodes 1 and 2 and the shortest path of the element in this case is not a geodesic. If the path is twice as long in the second case, if the kinetic energy of the element is the same as in case (I) and no potentials are present, then the time taken to cross between nodes 1 and 2 is twice as long. Therefore the action in case (II) is twice than the action in case (I). When we substitute these numbers in the expression for organization α (eq. 2), where n=1, one element, and m=1, one crossing between two nodes, then the denominator which is just the action of the element for that motion will be twice as large in the second case and therefore the result for the amount of organization is a half as compared to the first case.

## 4.2 Many elements and constraints

Now consider an arbitrary networks consisting of three, ten, thousands, millions and billions of nodes and edges, populated by as many elements and constraints, where the paths of the elements cross each other. The optimum of all of the constraints', nodes', edges' and elements' positions and the motions of the elements is the minimum possible action state of the entire system, providing a numerical measure for its organization. Notice that action is not at an absolute possible minimum in this case, but at a higher, optimal value. Action would be at its absolute minimum only in a system without any constraints on the motion of its elements, which is not the case in complex systems and networks. Nevertheless, action is at a minimum compared to what it will be for all other arrangements of nodes, elements and constraints in the system that are less organized. When we consider an open dynamical system, where the number and positions of nodes, edges, elements and constraints constantly changes, then this minimum action state is constantly recalculated by the system. It is an attractor state which drives the system to higher level of organization and this process can continue indefinitely, as long as the system exists. Achieving maximum organization is a dynamical process in open complex systems of constantly recalculating positions of nodes, edges, elements and constraints for a least action state and preserving those positions in a physical memory of the organization of the system.

Constraint minimization brings less action for the next cycle of motion of the elements along the same edge. There is less interaction with the constraint and less energy dissipation as a result. Action optimization brings the systems to a lower state of dissipation. If this process continues for a very long time, eventually all constraints will be minimized and there will be no energy dissipation. The system will reach a state of "Non-dissipative development". Non-dissipative systems abound. At certain temperatures, helium is superfluid, some materials are superconducting. At the temperatures of outer space, non-dissipative processes become more probable. But even at higher temperatures, all of the quantum processes are non-dissipative. There is no friction inside atoms. When systems are developed enough to continue their organization on quantum scale, they need not dissipate energy, therefore they may not require a constant source of energy. Those systems can continue self-organizing even in a universe, where all of the nuclear sources of energies in the stars have disappeared.

# 5 Exploring the limits for organization

## 5.1 An upper limit

The smallest possible discrete amount of action is one quantum of it, equal to the value of the Planck's constant. With self-organization the distances between the nodes shrink, the elements become smaller and the constraints for their motion decrease, for the purpose of decreasing of action (as in computer chips). The limit for this process of decrease of action is one quantum of it. If each motion uses the minimum of one quantum of action, then the value of the organization, α, is exactly one.

Can this value for organization be exceeded by a parallel processes, like quantum computing, where possibly with one, or a few quanta of action a vast number of computations can occur? Technically the crossing is still between two nodes, but it happens simultaneously along infinite number of different paths. It is like an infinite number of elements crossing between two nodes, each performing different computations. Alternatively, with decrease of the amount of action per crossing, it might be possible for the elements to cross several nodes (do several motions) with one quantum of action. In both of these cases the upper limit for organization, α, becomes very large and possibly infinity.

## 5.2 A lower limit

For a completely disorganized system, where the entropy is at a maximum, all points in the system are equally probable for an element to visit. In order to reach its final destination, an element of the system must visit all points in it (by definition for maximum entropy), thus maximizing its action for one crossing from any node 1 to any node 2. In this case, the action is extremely large and the organization, α, of this system is very close to zero.

Another way to describe the lower limit for organization of a system is when the constraint for the motion between nodes 1 and 2 is infinitely large and the path taken by the element to cross between the nodes is infinitely long. This also maximizes action and describes a completely disorganized system. The value for organization, α, in this case again approaches a limit of zero.

# 6 Mechanism of self-organization

When elements interact with constraints they apply force to minimize them, lowering their action for the next cycle. With the increase of quantity in a system, several elements can group on the same constraint to minimize it for less time. Decreased average action makes a system more stable, by lowering the energy needed for each motion. High average action, in disorganized system destabilizes it and above some limit it falls apart. Therefore a system with low enough average action can increase its quantity within limits of stability. Quantity and level of organization are proportional. If the quantity becomes constant, then the organization will reach a least action state and stop increasing. For continued self-organization an increase of the quantity is necessary. Quantity and level of organization of a system are in an accelerating positive feedback loop, ensuring unlimited increase of the level of organization in a system, unless it is destroyed by external influence, like limited resources, huge influx of energy, force impact, change in the conditions, etc.

**Further modification of the principle of least action for complex systems:** If there is strict time or energy constraint for a certain edge crossing, then the time and energy quantities in the action integral will attain different importance for that edge. In order to minimize action over its lifetime, over all edge crossings, the element may need to obey very strict time constraints for one edge. This means that the energy used may be a lot higher than if the time constraint was not there. The quantities in the action integral then need to be weighted differently, and the weight factor can be derived from the type and size of the particular constraint. This will modify the least action principle further and make it usable for complex systems.

**Cracking the chaos:** Cracks in rocks are energy spreading events in the least action way. The smaller the energy gradient with the surrounding media, the slower and more spherical the spread. The larger the energy gradient, the fractal dimensions change, the less populated the branching structure is. Electrical discharges and lightning are the same "cracks" in non-conducting air (large constraint for the flow of charge) and they form a flow fractal network which has the least action compared to all other possible networks. Dielectric fractal breakdowns form the same least action branching structures to propagate, spread and equilibrate energy. We can say that whichever is the fastest way to remove energy gradients it forms the flow network accommodating it. Cracks are flow networks and have the same branching morphology. Dendritic polymers make crystal crack in the isotropic amorphous surrounding. Bacterial colonies spread in fractal way, depending on the media surrounding them. Generalizing, "growth instabilities" spread as a fractal structure through interplay of the properties of the system and the surrounding media. In the same way organization spreads along a fractal flow network in the surrounding environment. Cities are flow networks growing in a fractal way. The central nucleus of the city discharges organization in unorganized area, and the least action way for that to happen is along a fractal flow network – cities are self-similar on different scales, small cities have the same fractal structure as large cities, growing cities keep their fractal structure. Cities are growth instabilities. New roads in unsettled areas are crack of organization into unorganized area and they originate new settlements as trees form new branches.

Cracks are energy penetration, by constraint minimization. The nonconductive environment is the constraint, and the flow network of energy (charge, water, blood, organization, etc. ) form paths, along which the action is minimized by the first elements passing for the following elements. This is the definition of a path, or road. So, this fractal network of paths is the least action way to spread the flow. It will take more energy for the elements to wander where there are paths, which will increase the action and decrease the organization of the system. The larger the difference with the surrounding area, the smaller number of paths, and the larger capacity along each path. Spreading homogeneously, in all directions, without paths maximizes the constraints that the elements encounter and minimizes the organization of the entire system. The fractal flow networks (discharges, cracks, roads) are the most organized state of the system, the least action, most efficient state. But the spreading and forming the network itself, comes from the fact that the energy tends to spread. It does not stay concentrated at one point. This is the maximum action principle, leading to the maximum entropy principle, causing the growth of systems. The systems achieve most action state, in the least action manner, with the least action flow network. With the fractal structure of a flow network, the elements meet the least amount of constraints – the Gauss least constraint principle.

If the energy amount at the center is increased, then the elements can overcome more constraints, new paths can be formed and the network will increase in density. Therefore the fractal dimension of the network will be proportional to the energy gradient from the center to the environment and the properties of the environment. If this difference is very low, it will spread in almost spherical manner. High difference will give rise to small number of branches with large capacity.

An amazing example is the comparison between the crack pattern in the glaze of a ceramic plate, which form in a hierarchical manner, first the largest and then the space between them is bridged with cracks, and the city street layout. If just the graphs of the two are compared, a difference cannot be found (P. Ball, "Branches" p.95, 96). This shows how much the self-organization in systems follows physical principles of efficiency and minimization which we think of as a sign of intelligence and intention. The city networks grow and then percolate, to connect the cities. This percolation (connection) transition, creates a global organized system. By growth, the maximum action and entropy principle, isolated self-organizing systems, percolate and connect into larger and larger systems until they occupy all of the available space in the universe.

**Matrix representation of self-organization of a complex system:** In mapping a complex system on a network, we need to know the action along each edge, which mathematically can be represented as a weighted adjacency matrix for the network. The time evolution of the network can be mathematically achieved by multiplying by a transform matrix that decreases the action for each edge, which physically can happen through distance, energy or time minimization. The transform matrix will take into account the processes of self-organization decreasing the constraints and one multiplication is one time step in the process. Then organization is calculated by finding the average action along all edges and the total number of elements and edge crossings that each of them does per unit time.

**Network representation of a complex system:**

| Network mapping | |
|---|---|
| **nodes** | endpoints |
| **edges** | trajectories |

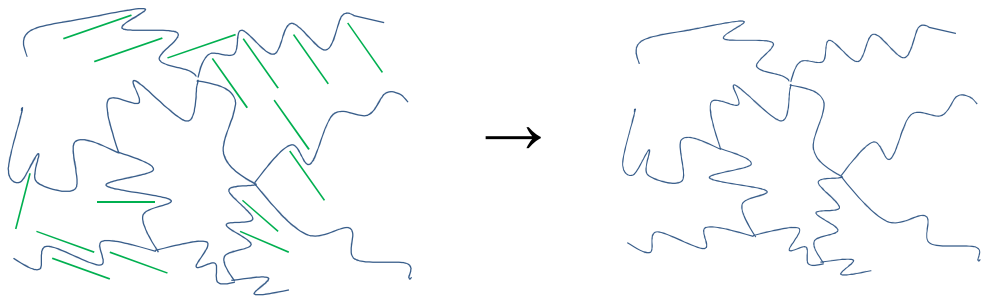

This figure represents a complex system as a general network, where the edges go around constraints (in green) which cause curvature. To the right the constraints have been removed to produce simpler equivalent graph, where the curvature (metric) of the edges reflects the presence of the constraints. **The paths are still geodesics in curved space - they are least action paths in it.**

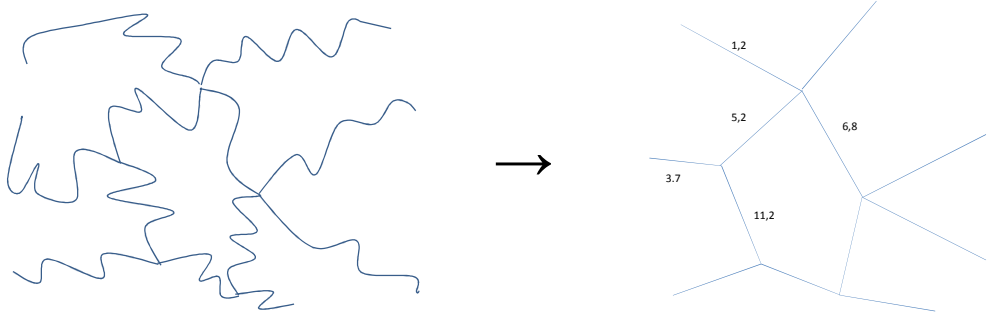

On this figure we transform the network to represent the length of the geodesic in curved space of each edge with a number.

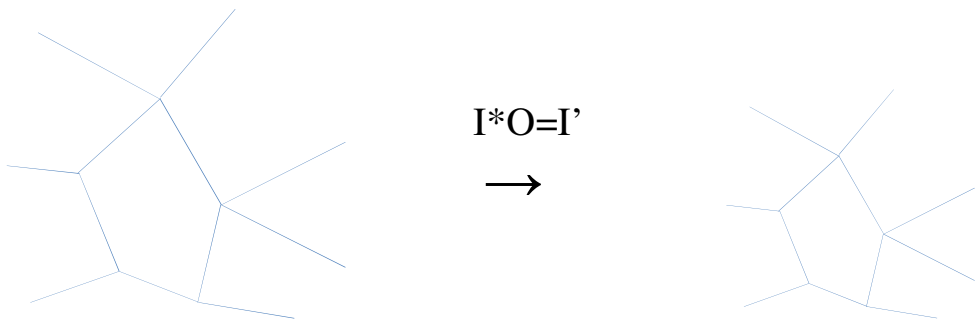

$I*O=I'$

Phase transition from long to short edges leading to reduction of action of elements. This is edges length minimization corresponding to increase of organization (action efficiency). Multiply the action matrix to organization O matrix, that reduces the values of I.

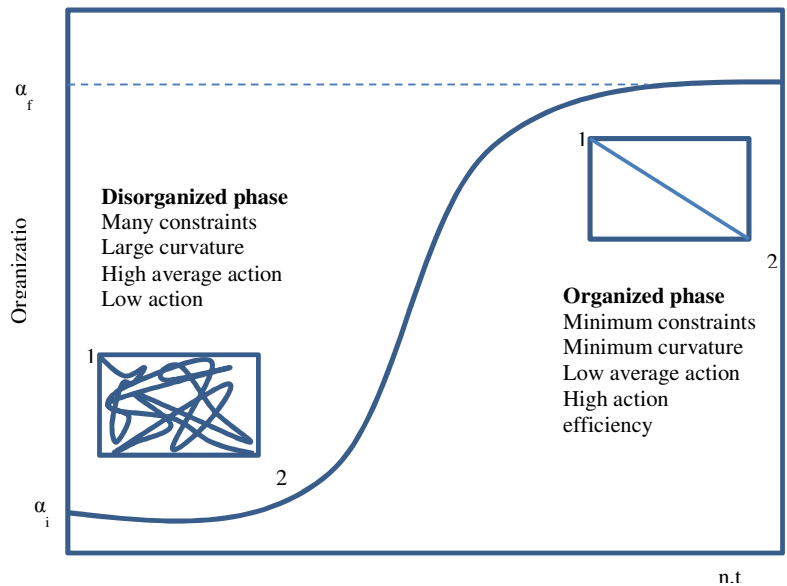

On the previous figure is represented the phase transition of decrease of the curvature of one edge of a complex system as a function of time and number of elements. According to the principle of least action, the action should decrease, and with time it becomes least by reorganizing the system.

**Procedure for calculating α:**
1. Map all of the points along the trajectories of the elements. Those are the edges of the network.
2. Find the length of the edges – the geodesic in this curved space by the constraints.
    a. This could be done directly, using the points.
    b. For more precise results, fit the points with the curve of best fit, and then use the equation of the curve to find its length, which is the distance between nodes – the length of the **geodesic edge in curved space**.
3. We can find closeness centrality of nodes or of the network in curved space. This will be **geodesic centrality in curved by constraints space**.
4. For each element in the system, for the particular edges that it traverses, find its action along those edges, m. Sum them. This is the total action for an element. Repeat the same for each of the elements. Sum the total actions for all elements, n. This is the denominator of α.
5. For large systems this may be computationally time consuming. For quick estimate a test element can be used. Its value for α of the system can be compared to the complete measurement of α to find the minimum number of edge crossings, m, that is necessary to bring the discrepancy with the exact calculation of α to acceptable level, which can be 1%, 0.01% or less, depending on the rigor needed.

**A network is** a supporting structure for a self-organizing complex system. It allows the flow of its elements and prevents jamming. But as such, the network itself is an expense for the system. Therefore the network needs to be minimized by the system, otherwise it will increase its action and therefore it will decrease its level of organization. So this is another "least" principle – what is the least possible network for the system that will allow the necessary amount of flow of elements at least action. A network larger than that will unnecessarily increase action. Examples are all networks in our society and in biology.

**Organization reactions:** In nuclear reactions adding new nucleons happens when an activation barrier is overcome and energy is lower in the bound state. In chemical reactions also the energy is lowered after a certain energy barrier is overcome. In networks, energy and time, i.e. certain amount of action is spent to create a new edge, which is analogous to a new bond between two nodes, but over the lifetime of the edge, it lowers the action in the network by shortening the geodesic distance. Action is spent to create a new edge, but more action is saved over the lifetime of that edge. The total action in the system is lowered. Creating a new edge is analogous to creating a new bond, and by lowering the action it increases organization. So we can call this an organization reaction. Organization reactions can also be of eliminating edges, if that lowers the action of the system. As an example, creating a new dendrite or axon in the brain is a new edge, which lowers the action of the neuronal network. Building a new bridge over a river, requires spending a lot of action, but over the lifetime of the bridge it will lower the action of the system.

**Growth in networks:** networks grow keeping their fractal structure, increasing the capacity of its edges, in the least possible way, with the growth of systems. The network is a mechanism that connects the parts in a whole, to move matter and energy from one specialized part to another. Growth is a fundamental problem, which should be considered in relation to the growth of the entire system. The network should be flexible to increase the number of nodes and edges, as well as the capacity of its edges. Chemical and nuclear reactions networks are the predecessors of the networks in higher order complex systems. For example, in stars, the number, and the connections of nuclear reactions with the evolution of the composition of a star, increases enormously. In chemistry, if we consider all of the reactions on our planet as a part of one huge chemical network, then with time the number of edges and nodes has been growing, leading to the complex molecules and of today.

**Organization phase transitions:** Organization and quantity are proportional. As it can be noted in most complex systems, quantity and therefore organization grows logistically, starting from one phase of organization, increasing exponentially and saturating at the new phase of organization. This logistic growth is characteristic of phase transitions in general. Therefore the organization phase transition can be described using the language of phase transitions in order to better understand it, and a phase diagram can be constructed to indicate the conditions at which transitions in organized systems occur. As in any other phase transition, if the control parameter is decreased, the phase transition may also reverse. So the rate of the phase transition, whether it will pause in the middle or reverse, will depend on the change in the control parameter. The "average action efficiency phase transition" or for short "organization transition" is a transition from state with high average action to state with low average action, from low action efficiency to high action efficiency, from state with low organization to state with high organization, from disorganized to organized phase of a complex system. In general it is series of phase transitions, each of which is a transition from one level of organization to a higher level of organization. This transition does not exist for free elements, outside complex system, because they always obey the principle of least action and at maximum efficiency. Only in complex systems, where action is increased by constraints, reorganization can lead to intermediate states of action.

**Maximum total action and minimum unit action in complex flow networks:**

**Maximum total action** is the tendency of developing and self-organizing systems to expand and to increase their quantity: number of elements and constraints, matter, energy and space occupied, to the longest time possible, which increases the total amount of action as the sum of the actions of all of its elements. The evidence for that is in the fact that all self-organizing systems, which are developmental and evolutionary by their nature, tend to occupy as much of the surrounding resources as possible. Examples include all biological species and human societies. Those systems always grow to the maximum limits imposed by their environment in the form of constraints that cannot yet be overcome by those systems. In the process systems compete, and the less organized ones may actually shrink and die off, but nevertheless the tendency is the same. This is another attractor for systems, but how close are they to this attractor depends on the interplay between their organization, the environmental constraints and the competing systems, which can also be counted as environmental constraints. This is the expansionist quality of self-organizing systems, which is connected to the **least unit action** per one edge crossing of one element, which defines the internal organization of the system. The lower the unit action in the system, the more successful it will be in expanding. This tendency, to minimize the unit action, in order to maximize the total action and spread the organization that provides least unit action over larger amount of matter is the driving force behind self-organization. On the other hand, the larger the system, the more grouping of elements can occur, the faster the constraints can be minimized and the greater the organization becomes. The maximum total (quantity) and the minimum unit action (quality) are in positive feedback mechanism driving the development of complex systems. The increase of the total action necessitates and accelerates the decrease of unit action, and the decrease of unit action allows increase of the total action in the system by being able to overcome more external constraints and expand. Even though both of these principles seem equally important, it may be the maximum total action that is the leading driver of development of complex systems, and the least unit action a tool for that expansion. The quality serves to increase the quantity. Examples abound: when a system is surrounded by insurmountable constraints, such as the isolation of Australia from the rest of the continents, then the system that exists there cannot expand anymore and achieves a local minimum of action, and stops reorganizing (decreasing of unit action) until that continent is joined with others. Countries that use isolationist policies also limit themselves to a smaller size of their system and achieve a local minimum that is destroyed when the country opens to the rest of the world. The larger a system is, the lower the unit action it can achieve. The lower the unit action, the more it can expand. **Growth** in systems is the most readily observed and the leading phenomenon behind self-organization. I want to note here, that this line of thinking is for complex, self-organizing systems, which can be represented by flow networks, persisting and increasing their level of organization with time. Well known non-equilibrium thermodynamics phenomena, such as dissipative structures, sand pile models, whirlpools, and other stable over certain amount of time structures, may be relevant of some of the mechanisms and aspects of developing systems, but are not developing themselves, i.e. they do not continuously increase their level of organization and do not expand driven by internal mechanisms. Rather, they are a temporary response in a highly non-equilibrium situation, minimizing the action of the energy flow, but they disappear when the system is equilibrated. They do not have a character of a network of interacting elements moving in cycles, interacting with constraints and decreasing the unit action for their edge crossings at each cycle. They are temporary least action structures, same as a falling object in gravity is a temporary situation until its driving potential energy are equilibrated, and there is no further cause for motion. Developing complex systems actively seek and create non-equilibrium energy settings where the potential differences create the forces that move their elements along the network (in following sections).

**Maximum action** provides all of the redundancy and robustness in a system necessary to increase its probability for survival in time. The tendency for a system to spread is the same as the tendency of a molecule to visit all possible states in a container, or this is the tendency to increase entropy by visiting the states in the surrounding environment, in order to equilibrate the system with the rest of space. This is a statement of the second law of thermodynamics, and the tendency to do this as fast as possible, is a statement of the Maximum Entropy production Principle.

**Attractive and repulsive potentials in nature and their role in self-organization:** Temperature differences provide repulsive potentials, where particles move from high to low potential, from high to low energy region, obeying the principle of least action, eventually equilibrating the two. This is a statement of the Maximum Entropy Production Principle (MEPP) for systems far from equilibrium. Those dispersive potentials create dissipative structures. If we consider energy gradients as the motive forces of elements in complex networks, then we can define the nodes as places of different potential. The motions of the elements under the influence of these potentials are caused by those energy gradients which define forces on the elements. This is the cause for motion of elements in complex networks, and in general in complex systems. Using this reasoning, the nodes and edges in those networks become well defined.

Is this repulsive potential the only force that exists in nature moving elements in complex systems or creating the conditions for those motions? The answer is "no", because we are ignoring all other forces and potentials, which ultimately can lead to a more complete explanation of the existence of dissipative structures and complex systems in general. All those other potentials are well known in Physics, and need to be included in the framework of complexity theory. Any existing potential difference provides motion force for the elements along the edges, defines start and endpoints and if it is ignored, our understanding of complex networks cannot be complete. Other potentials existing in nature may be attractive, which means that they are not dissipative but aggregative. Their operation does not lead to their decrease in time, as the energy or temperature repulsive gradients do, but to their increase. Instead of dispersing, they are collecting the elements on which they act and increase their concentration. Instead of leading to equilibrium and death of the system, they lead to disequilibrium and increase the rate of the processes, therefore allowing higher levels of organization.

One such potential, which is very weak and does not directly act in complex systems, but on a cosmic scale is gravity. This gradient is directly responsible for the available matter concentration and for the energy and temperature gradients that move the elements in complex systems. Gravitational potential, increases as more elements are aggregated. It also concentrates energy in stars and planets, heating them and providing the energy gradients leading to dissipation obeying MEPP. Without considering this attractive potential, there is no explanation for why matter is available and concentrated enough for complex systems to form and for the dissipative energy gradients moving the elements along the edges. If an attractive potential did not operate first to create regions with different concentration of matter and different temperature, nodes in complex systems would have been undefined – all points would have been at equilibrium, at the same potential, therefore there would have been no motive force for the elements to cross the edges. The initial and starting points for the motions would have not being defined. Therefore gravity, as an attractive potential, in positive feedback with its operation (the gravitational potential difference between two points increases as gravity attracts matter), is the precursor for formation of complex systems. It provides the high concentration of matter necessary for dissipative structures and the energy gradients to define all of the processes occurring in them. Without the existence of this attractive potential, the universe would have been homogeneously seeded with atoms at the same temperature, at extremely low density and the MEPP principle would have been mute, since the entropy would always stay at a maximum. No forces would have existed to move any of the elements. This would have been the heath death of the universe as predicted by the second law of thermodynamics. This scenario was prevented by the gravitational potential, collecting matter in stars and planets and creating differences in density and energy, which then are used to form functioning complex systems.

We can treat the attractive potential of gravity as initial cause for MEPP and for complex systems. We cannot ignore its continuing action in the universe, new stars formation and the resulting energy flows, and especially the synthesis of heavier chemical elements, without which complex systems would not have existed as well. We have to place gravity in the origin of chain of events and principles leading to the complex systems that we observe today. Are there other potential gradients that move elements in complex systems? Several other forces, electric, magnetic, strong and weak nuclear, are responsible for gradients collecting, not dispersing matter on a smaller scale and building the blocks for complex systems. Without the last, there would have been no atoms and molecules on which gravity to act. Those, precede even the attractive action of gravity.

Non-dissipative potentials, such as the above mentioned, are critical in collecting energy into matter, building molecules and cells and allowing the complex systems to exist and function. Those potentials act in the everyday operation of complex systems, preventing many of the effects of energy dissipation, which corresponds to Prigogine's Least Entropy Production Principle (LEPP). For example, without the attractive electrostatic potential, the electrons would have been dissipated from their nuclei almost at zero temperature, rendering atoms nonexistent. The same forces keep molecules together at temperatures necessary for operation of complex systems, preventing dissipative processes. Electrostatic gradients are the ones moving electrons in computer circuits and along the internet, not temperature gradients. How can we understand functioning of complex systems not including them in the framework? Neurons in the human body send signals with the same electrostatic potentials, but are not moved by temperature differences and therefore are not a subject to the MEPP, even though they are connected to the energy dissipation by the organism.

Cells aggregate in multicellular organisms, not only by entropy production, but also under aggregative action efficiency difference. Similar attractive, not dissipative differences, aggregate people in cities. The action efficiency difference between cities and remote regions attract people. They provide an aggregative potential. As important are the MEPP, and the Prigogine's LEPP, which are both driven by dissipative, repulsive potentials for understanding complex systems, the same level of consideration should be given to the attractive, non-dissipative potentials, which create the energy difference necessary for dissipation to exist and move the elements of complex systems by other potential differences.

Repulsive potential gradients are dissipative, dispersive, they increase symmetry and decrease differentiation. Attractive potential gradients are aggregative, accumulating and they decrease symmetry and increase differentiation. Repulsive forces decrease their potential difference therefore the energy available to do work and slow down the system in a negative feedback mechanism. Attractive forces increase their potential difference, therefore the available energy to do work and speed up the system in a positive feedback mechanism.

## 7 Conclusions

The principle of least action for a networked complex system (eq. 1) drives self-organization in complex systems and the average action is the measure of degree to which they approach this least action state. Actions that are less than their alternatives are self-selected. Progressive development, as self-organization, is a process of minimization of action. In open systems there is a constant change of the number of elements, constraints and energy of the system and the least action state is different in each moment. The process of self-organization of energy, particles, atoms, molecules, organisms, to the today's society is a process of achieving a lower action state, with the least action as a final state. The laws of achieving this least action state are the laws of self-organization. The least possible action state is the limit for organization when time is infinite and all elements in the universe are included.

The state of nodes, edges, constraints and elements that determines the action for one motion in a system is its organization. With its measure α (eq. 2) we can compare any two systems of any size and the same system at two stages of its development. It distinguishes between systems with two different levels of organization and rates of self-organization and is normalized for their size. The measure can be applied to all systems and researchers in all areas studying complex systems can benefit from it. With a quantitative measure we can conduct exact scientific research on self-organization of complex systems and networks, progressive development, evolution and co-evolution, complexity, etc.

## 8 Future work

If a system does not grow in quantity, it does not need greater action efficiency - quality. A single cell does need a blood flow network, but a large amount of cells living together does. If people live in a small autonomous village, they do not need any other communication tool than their voices. But if their system grows, they may need mail, telegraph, telephones and eventually internet. For all practical purposes, now our communications are at the speed of light. If the speed was faster, it will not have affected much our lives, except for stock trades. But, if we become a civilization on the scale of the solar system, or at even larger scale, the speed of light is too slow for communication. Then we can have benefit if we use faster communications. It is not guaranteed that we will find them, but at least this line of logic explains why with greater size of a system (greater quantity), more efficiency (quality) can be utilized and possibly achieved. The larger quantity increases the need of quality. Otherwise the system will fall apart, as a galactic civilization that uses communications at the speed of light. If it falls apart, then it cannot achieve the same rate of increase of its organization, as if it was intact, because the sizes of the separate systems will be much smaller than the size of the single galactic system. We arrive at a mathematical proportionality, which is that the unit action efficiency is inversely proportional to the total amount of action in a system, with a constant of proportionality, let's call it k. This is the same statement as in the law of quality-quantity transition. The exact value of this constant of proportionality is hard to determine. For sure the relation is statistical – there will be deviations in the proportionality values in different systems. Then there is the nonlinear logistic growth relation and the punctuated equilibrium stepwise growth curve, so k cannot possibly be a constant. It has some range. What is it? We know something about the energy efficiency in animals, which is the Kleiber's law that the energy expenditure increases with a power of ¾ as a function of mass. But we do not know how the time efficiency changes, in order to determine the action efficiency of those organisms. Does the efficiency change in the same way in other systems? At this time we do not have enough data to answer this question, which makes it a great research question: **what is k?** What is it for different systems? What is its nonlinear, statistical, probabilistic nature? There is so much to be learned in just answering this one specific question.

**Acknowledgments.** The author thanks Assumption College for support and encouragement of this research, and Prof. Slavkovsky, Prof. Schandel, Prof. Sholes, Prof. Theroux and Prof. Cromarty for discussion of the manuscript. The author thanks Francis M. Lazarus, Eric Chaisson, John Smart, Clement Vidal, Arto Annila, Tessaleno Devezas, Yaneer Bar-Yam and Atanu Chatterjee for invaluable discussions.